\documentclass[prl,aps,superscriptaddress,twocolumn,floatfix,10pt]{revtex4-2}

\usepackage{ragged2e}

\usepackage{dsfont}

\usepackage{mathtools}
\usepackage{amsmath}
\usepackage{amssymb,mathrsfs}
\usepackage{epstopdf}
\usepackage{mathtext}
\usepackage{yfonts}
\usepackage{bm}

\usepackage{soul}

\usepackage{color}
\usepackage[colorlinks=true,linkcolor=blue]{hyperref}

\newcommand{\br}{{\bm r}}
\newcommand{\bbeta}{{\bm \beta}}

\usepackage{xcolor}
\usepackage{graphicx}% Include figure files
\usepackage{dcolumn}% Align table columns on decimal point
\usepackage{bm}% bold math
\usepackage[caption=false]{subfig}
\usepackage{physics}

\begin{document}

\preprint{APS/123-QED}

\title{Observation of linear and nonlinear light trapping on topological dislocations}

%\author{Autor1}
% \affiliation{Departamento de Física Teórica e Experimental, Universidade Federal do Rio Grande do Norte,
%59072-970 Natal, Rio Grande do Norte, Brazil}
%\affiliation{Université Côte d’Azur, CNRS, Institut de Physique de Nice, 06560 Valbonne, France}

%%%%%%%%%%%%%%%%%%%%%%%%%%%%%%%%%%%%%%%%%%%%%%%%%%%%%%%%%%%%%%%%%%
\author{S.~K.~Ivanov}
\email[Correspondence email address: ]{sergei.ivanov@uv.es}%
\affiliation{Instituto de Ciencia de los Materiales, Universidad de Valencia, Catedr\'{a}tico J. Beltr\'{a}n, 2, 46980, Paterna, Spain}

\author{A.~V.~Kireev}
\affiliation{Institute of Spectroscopy, Russian Academy of Sciences, 108840, Troitsk, Moscow, Russia}
\affiliation{Moscow Institute of Physics and Technology (National Research University), Dolgoprudny, 141701, Russia}

\author{K.~Sabour}
\affiliation{Moscow Institute of Physics and Technology (National Research University), Dolgoprudny, 141701, Russia}

\author{N.~S.~Kostyuchenko}
\affiliation{Institute of Spectroscopy, Russian Academy of Sciences, 108840, Troitsk, Moscow, Russia}
\affiliation{Quantum Technology Centre, Faculty of Physics, M. V. Lomonosov Moscow State University, 119991, Moscow, Russia}

\author{S.~A.~Zhuravitskii}
\affiliation{Institute of Spectroscopy, Russian Academy of Sciences, 108840, Troitsk, Moscow, Russia}
\affiliation{Quantum Technology Centre, Faculty of Physics, M. V. Lomonosov Moscow State University, 119991, Moscow, Russia}

\author{N.~N.~Skryabin}
%\affiliation{Institute of Spectroscopy, Russian Academy of Sciences, 108840, Troitsk, Moscow, Russia}
\affiliation{Quantum Technology Centre, Faculty of Physics, M. V. Lomonosov Moscow State University, 119991, Moscow, Russia}

\author{I.~V.~Dyakonov}
\affiliation{Quantum Technology Centre, Faculty of Physics, M. V. Lomonosov Moscow State University, 119991, Moscow, Russia}

\author{A.~A.~Kalinkin}
\affiliation{Quantum Technology Centre, Faculty of Physics, M. V. Lomonosov Moscow State University, 119991, Moscow, Russia}

\author{V.~O.~Kompanets}
\affiliation{Institute of Spectroscopy, Russian Academy of Sciences, 108840, Troitsk, Moscow, Russia}

\author{S.~P.~Kulik}
\affiliation{Institute of Spectroscopy, Russian Academy of Sciences, 108840, Troitsk, Moscow, Russia}
\affiliation{Quantum Technology Centre, Faculty of Physics, M. V. Lomonosov Moscow State University, 119991, Moscow, Russia}

\author{S.~V.~Chekalin}
\affiliation{Institute of Spectroscopy, Russian Academy of Sciences, 108840, Troitsk, Moscow, Russia}

\author{A.~Ferrando}
\affiliation{Instituto de Ciencia de los Materiales, Universidad de Valencia, Catedr\'{a}tico J. Beltr\'{a}n, 2, 46980, Paterna, Spain}

\author{V.~N.~Zadkov}
\affiliation{Institute of Spectroscopy, Russian Academy of Sciences, 108840, Troitsk, Moscow, Russia}
\affiliation{Faculty of Physics, Higher School of Economics, 105066, Moscow, Russia}

\author{Y.~V.~Kartashov}
%\email[Correspondence email address: ]{kartashov@isan.tro\-itsk.ru }%
\affiliation{Institute of Spectroscopy, Russian Academy of Sciences, 108840, Troitsk, Moscow, Russia}
%%%%%%%%%%%%%%%%%%%%%%%%%%%%%%%%%%%%%%%%%%%%%%%%%%%%%%%%%%%%%%%%%%

\date{\today}

\begin{abstract}

Topological dislocations in otherwise periodic lattices represent global structural defects that, nevertheless, typically leave the lattice periodicity intact far from the dislocation. Such dislocations arise in diverse physical systems ranging from crystalline solids, acoustic and photonic lattices and crystals to matter waves in optical lattices. Dislocations drastically affect the evolution of wave excitations in their vicinity, enabling novel mechanisms for trapping on topological defects and controlling the energy flow. Moreover, when combined with nonlinearity, such systems give rise to new types of self-sustained states of topological origin that have never been observed to date. Here we demonstrate experimentally, for the first time at optical frequencies, the waveguiding at various types of topological edge dislocations, resulting in the formation of localized photonic eigenstates with distinct and tunable shapes. Using femtosecond laser-writing techniques, we fabricated waveguide arrays with precisely tailored dislocation parameters, enabling full control over the degree of localization and internal structure of the associated modes. We further demonstrate both theoretically and experimentally that in the high-power regime, the families of thresholdless dislocation solitons bifurcate from such modes, which inherit shape diversity of their linear counterparts. Our results reveal a nontrivial interplay between nonlinearity and global lattice deformations and establish dislocation solitons as a new class of nonlinear topological states. They may stimulate the observation of new types of nonlinear states and interaction scenarios for excitations in nonlinear physical systems, where lattices with controllable global deformations can be created.

\end{abstract}

\maketitle
%%%%%%%%%%%%%%%%%%%%%%%%%%%%%%%%%%%%%%%%%%%%%%%%%%%%%%%%%%%%%%%%%%

% original text

%\section{Introduction}\label{sec1}

The concept of dislocations dates back to studies of plastic deformation of metals in the 19th century, although it was formally established only in the 1930s, when the topological nature of dislocations was proposed~\cite{Burgers-1939}. In crystalline solids, dislocations are now recognized as ubiquitous structural defects that play a pivotal role in shaping not only mechanical, but also electronic and optical properties of materials, including semiconductors and ionic crystals~\cite{Hull-2011, Butz-2014}. At their core, dislocations represent localized mismatches in the lattice structure, that can be classified into two primary types. Edge dislocations -- which are the focus of the present work (in the optical context) -- correspond to the termination (or insertion) of an atomic plane within a crystal or to the merging of several planes, resulting in a lattice mismatch across a glide plane. Screw dislocations, on the other hand, produce a helical distortion around the cut through the crystal that resembles a spiral staircase for atomic planes. Dislocations can also be of mixed type. Each dislocation is characterized by a dislocation line and a Burgers vector, which quantifies the lattice mismatch encountered when encircling the defect.

Dislocations attract considerable attention due to their global, topological properties~\cite{Mermin-1979, Kleman-2008}. Dislocations represent a real-space topological defect, characterized by a conserved Burgers vector, which makes them a singular feature of the lattice that cannot be removed by smooth deformations. Importantly, there exists a connection between real-space and momentum-space topologies, as dislocations have been shown to probe the band topology, and give rise to protected modes whose existence can be explained via the bulk-boundary correspondence~\cite{Hasan-2010, Qi-2011, Ozawa-2019, Xie-2021}. This has revealed the connection of lattices with dislocations to topological phases of matter, including crystalline topological insulators~\cite{Ran-2009, Teo-2010, Juricic-2012, Slager-2014, Miert-2018}
%\cite{Ran-2009,Teo-2010,Imura-2011,Juricic-2012,Slager-2014,Teo-2017,Miert-2018}
and topological semimetals~\cite{Juan-2014, Chernodub-2017, SotoGarrido-2020}.
%\cite{Juan-2014,Chernodub-2017,Sumiyoshi-2016,SotoGarrido-2020}
Localized states of topological origin emerging on dislocations have been explored for a broad range of linear platforms, from acoustic~\cite{Ye-2022}, to
%\cite{Xue-2021,Lin-2022,Ye-2022,Zhou-2025}
mechanical~\cite{Paulose-2015},
%\cite{Paulose-2015,Grinberg-2020}
electronic~\cite{Hughes-2014, Hamasaki-2017},
%\cite{Hughes-2014,Hamasaki-2017,Nayak-2019}
and ultracold atomic systems~\cite{Zhang-2015}. Dislocations may also be nested and result in the appearance of new states in materials in more exotic phases, such as in higher-order topological insulators~\cite{Queiroz-2019, Roy-2021, Yamada-2022}, {as well as in systems featuring the non-Hermitian skin effect~\cite{Schindler-2021,Bhargava-2021,Panigrahi-2021,Chadha-2021,Wu-2025}.}

Topological defects may also be created in photonic systems, providing rich opportunities for precise control of global symmetry and local lattice structure (see also recent reviews~\cite{Lin-2023}). For example, photonic Floquet systems, which change periodically in the evolution variable, have been used to realize screw dislocations~\cite{Bi-2017,Nag-2021} that can support~\cite{Lustig-2022} robust one-dimensional gapless modes in three-dimensional settings. At the same time, topological modes at edge dislocations -- which are the focus of this work -- have never been observed at optical frequencies and have only been considered in structures based on photonic crystals in the microwave range~\cite{Li-2018, Liu-2024}. {For example, very recently, selective trapping of excitations in the microwave frequency range has been demonstrated in photonic crystal microcavities featuring two separated topological dislocations \cite{Liang-2025}.} It should also be mentioned that in addition to dislocations, a different type of disclination defects can be created based on an originally periodic structure that can host modes of topological origin in higher-order topological insulator phases~\cite{Peterson-2021, Liu-2021, Wang-2020, Ren-2023}. Modern technologies allow the creation of highly controllable waveguiding structures in transparent nonlinear optical materials that mimic two-dimensional crystalline materials and realize lattices with different types of edge dislocations. This motivates the exploration of the impact of topological defects on modal structure of the system and observation of new mechanisms of topological origin for light confinement and control at optical frequencies. Such studies may unlock fundamentally new ways to manipulate light via the real-space topology.

Moreover, optical materials provide a unique platform for the exploration of the interplay between topological trapping and nonlinearity, owing to their pronounced nonlinear response. In conventional periodic lattices, the interplay between nonlinearity and diffraction in a periodic refractive index landscape results in the formation of lattice solitons ~\cite{Lederer-2008, Chen-2012, Kartashov-2019}, whose characteristic feature in multidimensional settings is the presence of a power threshold required for soliton formation. The spectrum of linear lattice appears to be a key factor determining the properties and stability of such self-sustained states in both periodic \cite{Fleischer-2003, Efremidis-2003} and aperiodic lattices~\cite{Freedman-2010, Ablowitz-2012, Fu-2020}. While the evolution of matter-wave solitons around dislocations nested in quasi-one-dimensional lattices~\cite{Kartashov-2005, Kartashov-2006} and optical solitons in two-dimensional lattices with edge dislocations~\cite{Ablowitz-2006,Bagci-2014} were studied theoretically, they have never been observed experimentally. Given that the properties of solitons are closely tied to the underlying linear lattice spectrum, which may change qualitatively depending on the dislocation type and geometry, the experimental exploration of such novel self-sustained states is of considerable interest. Notice that they are principally different from lattice solitons, as they cause local (rather than global) deformations in the lattice~\cite{Martin-2004}.

In this paper, we present the first experimental observation at optical frequencies of localized linear modes with distinct internal structure bound to the edge dislocations and, crucially, their nonlinear counterparts -- dislocation solitons. For our experiments, we have fabricated waveguide arrays in fused silica via femtosecond laser direct writing, enabling precise control over the dislocation geometry. We find that the dislocation configuration strongly influences the degree and nature of localization, giving rise to a diverse set of linear dislocation modes and solitons bifurcating from them. We identify the families of dislocation solitons that can be entirely stable and thresholdless -- an essential distinction from solitons in periodic lattices. Experimentally, we observe a clear transition from linear dislocation states to solitons as the input pulse power increases. These findings establish dislocations in the lattice structure as a powerful mechanism for control of light localization, bridging the gap between real-space topology and nonlinear photonics.

{It is important to stress that from the point of view of study of self-action effects on dislocations, there exist qualitative differences between formally unbound systems based on waveguide arrays, where light propagates along the $z$ direction, in which the length of the waveguides can be arbitrary, and various microcavity systems, such as photonic crystal microcavities or polariton microcavities \cite{ Li-2018, Liu-2024, Liang-2025}, where the field is confined within the microcavity due to reflection from top and bottom layers defining microcavity. The most substantial difference is that waveguiding system considered here is characterized by the low refractive index contrast, hence nonlinear contribution to the refractive index due to Kerr nonlinearity in it may be comparable with refractive index modulation depth that opens the possibility for dramatic reshaping of excited nonlinear dislocation states upon increase of power of laser radiation. In contrast, photonic crystal microcavities are characterized by large refractive index contrast, and even though strong mode confinement substantially enhanced nonlinear effects in them, nonlinearities in such systems are usually considered in perturbative regime, when they can shift corresponding frequencies, but weakly affect mode profiles at accessible power levels. Another substantial difference is that photonic crystal and polariton microcavities are essentially lossy structures due to mode leakage and intrinsic absorption of corresponding materials, hence excitation of stationary states in them requires constant pumping. In combination with nonlinear response and losses, such pumping may give rise to tilt of the resonances arising around eigenfrequencies of modes of the system and even to bistability effects. In contrast, waveguide arrays inscribed in transparent dielectric material considered here are characterized by very low absorption levels, hence excited nonlinear states experience very weak attenuation, do not show bistability and do not possess internal currents always existing in dissipative microcavities. The type of the excited nonlinear state is determined exclusively by the profile of the input beam, but not by its frequency, offering considerable flexibility in selective excitation of different nonlinear states that may coexist in our system without coupling with unwanted and spatially considerably separated modes. Finally, in study of self-action phenomena in microcavities, such as optical bistability and nonlinear nonreciprocity, it has been preferable to employ only a few nonlinear elements, in order to avoid issues related to multistability and dynamical instabilities. In contrast, waveguide arrays with dislocations allow to consider the systems with unconstrained size, where stability properties of solitons do not change once the system size exceeds certain minimum.}

%%%%%%%%%%%%%%%%%%%%%%%%%%%%%%%%%%%%%%%%%%%%%%%%%%%%%

\section*{Results}\label{sec1}

%%%%%%%%%%%%%%%%%%%%%%%%%%%%%%%%%%%%%%%%%%%%%%%%%%%%%%%%%%%%%%%%
\begin{figure*}[t]
\centering
\includegraphics[width=1.00\linewidth]{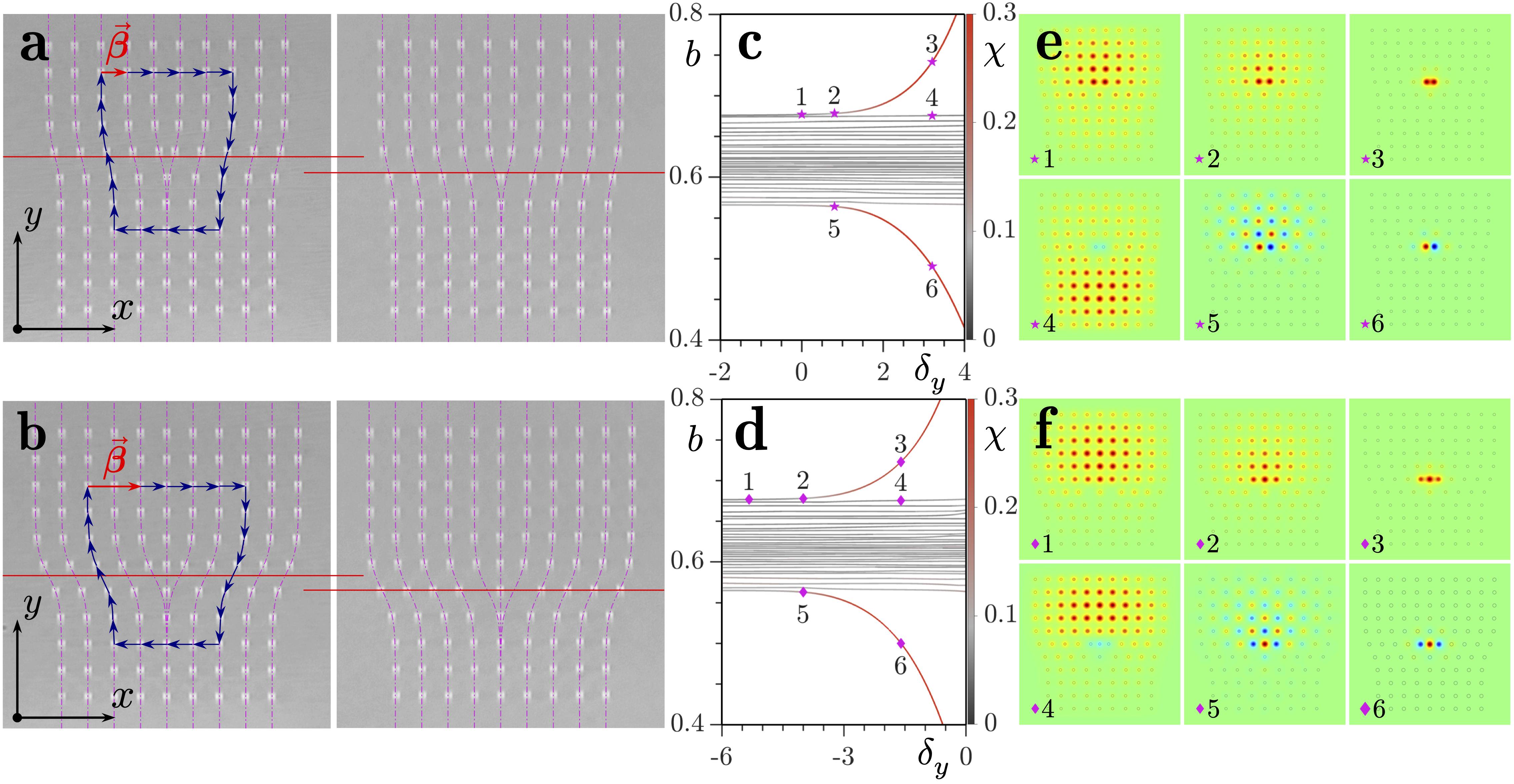}
\caption{{\bf Eigenmodes of arrays with dislocation.} {\bf (a–b)} Microphotographs of the arrays with dislocations, where either two {\bf (a)} or three {\bf (b)} waveguide layers merge into one. The red
horizontal lines are drawn in accordance with the value of the shift parameter $\delta_y$, which is larger in the left panels. Magenta dashed-dotted lines trace the waveguide positions within the arrays. The red arrows correspond to the Burgers vectors $\bbeta$ {\bf (c–d)} Linear spectra of the dislocation arrays as a function of $\delta_y$ parameter for configurations with two {\bf (c)} and three {\bf (d)} merging layers. The color of the curves in {\bf (c-d)} illustrates the form-factor of the eigenstates that quantifies their localization degree (red color corresponds to well-localized modes, while dark gray corresponds to extended states). Points marked with magenta stars in {\bf (c)} correspond to eigenmodes shown in {\bf (e)}, while points marked with magenta diamonds in {\bf (d)} correspond to eigenmodes in {\bf (f)}. Black circles in {\bf (e-f)} indicate waveguide positions. Arrays and eigenmodes are displayed within the window $x,\,y\in[-250\,\mu \textrm{m},+250\,\mu \textrm{m}]$. Here and below $p=3.1$, $d=4.0$, $w_x=0.75$, and $w_y=0.65$.}
\label{fig1}
\end{figure*}
%%%%%%%%%%%%%%%%%%%%%%%%%%%%%%%%%%%%%%%%%%%%%%%%%%%%%%%%%%%%%%%%

\subsection*{Lattices with edge dislocations}\label{subsec01}

To achieve light trapping on topological lattice defects, we create several distinct types of lattice edge dislocations. As a base structure, we use a standard periodic square lattice (realized as an array of fs-laser written waveguides) with period $d$, which is known to support only extended linear Bloch waves. The introduction of an edge dislocation modifies the real-space topology of the lattice, enabling the emergence of localized modes even in the linear case. It turns out that the nature of the introduced dislocation strongly impacts the spatial profile and localization properties of the emerging eigenmodes. Figures~\ref{fig1}(a) and~(b) illustrate microphotographs of two representative fs-laser written waveguide arrays with edge dislocations, where two or three waveguide layers (indicated with magenta dashed-dotted lines) merge into one, respectively. We have also analyzed other types of edge dislocations, including configurations with the merging of four layers and with layer termination, as presented in the \textbf{Supplemental Materials}. In all cases, the dislocation is characterized by a Burgers vector $\bbeta$, which quantifies the lattice distortion. To define $\bbeta$, we construct a closed contour in the initial unperturbed lattice, which, however, fails to close after introduction of the edge dislocation with center inside the contour [see blue arrows in Figs.~\ref{fig1}(a) and~(b)]. The additional vector that is required to close the contour corresponds to the Burgers vector, which is shown by the red arrow in the figures above. In the case of the merger of two layers, $|\bbeta| = d$, whereas in the case of three merging layers, $|\bbeta| = 2d$. This difference leads to a qualitatively distinct spectrum of localized modes, as shown below. Specifically, for $|\bbeta| = d$, the linear modes are predominantly confined to two waveguides near the center of the dislocation, while for $|\bbeta| = 2d$, they predominantly span over three waveguides. {Increase of the number of merging layers producing structures with larger $|\bbeta|$ typically leads to the expansion of localized modes along the red line in Figs.~\ref{fig1}(a) and~(b), but the very fact of localization will persist, as demonstrated in the \textbf{Supplemental Materials}. This also enriches the variety of eigenmodes that can be supported by dislocation, because eigenmodes occupying different number of waveguides become possible. The structure of such modes, as well as the point, where they split from the bulk band are determined by the particular configuration of waveguide layers around the dislocation, as shown below.} Surprisingly, we found that the degree of localization of modes emerging at the dislocation is highly sensitive to the precise geometry of the lattice around the dislocation center. By varying the relative positions of nearby waveguides, one can tune the dislocation modes from strongly localized to fully delocalized despite the fact that the topological defect is always present in the lattice.

%%%%%%%%%%%%%%%%%%%%%%%%%%%%%%%%%%%%%%%%%%%%%%%%%%%%%%%%%%%%%%%%
\subsection*{Linear spectrum and solitons at the dislocation}\label{subsec11}

To describe unusual features of light localization at topological dislocations, we consider the propagation of paraxial light beams in our shallow waveguide arrays governed by the nonlinear Schr\"{o}dinger equation for the dimensionless light field amplitude $\psi$:
\begin{align}
\label{NLS}
    i\frac{\partial\psi}{\partial z} = -\frac{1}{2}\left(\frac{\partial^2 \psi}{\partial x^2} + \frac{\partial^2 \psi}{\partial y^2}\right) -\mathcal{R}(x,y)\psi - |\psi|^2\psi.
\end{align}
Here, $x$ and $y$ denote the dimensionless transverse coordinates, while $z$ corresponds to the dimensionless propagation distance. The waveguide array featuring a dislocation is described by the function $\mathcal{R}(\br) = p\sum_{n,m} e^{-\left[(x-x_{n,m})^2/w_x^2+(y-y_{n,m})^2/w_y^2\right]}$, where $(x_{n,m},\,y_{n,m})$ stand for the coordinates of waveguide centers, while waveguides feature Gaussian shapes. The distance between waveguides far away from a dislocation is $d=4$, and $w_x=0.75$, $w_y=0.65$ are the widths of each waveguide along the $x$ and $y$ axes (they are slightly elliptical due to the employed fs-laser writing method \cite{Lustig-2022, Ren-2023, Skryabin-2024}). The depth of the waveguides $p = 3.1$ at the working wavelength $\lambda=800~\textrm{nm}$ corresponds to the refractive index contrast $\delta n\approx3.5\cdot10^{-4}$. The physical $10~\textrm{cm}$ length of the sample corresponds to the dimensionless propagation distance of $z\approx 88$. We also account for the focusing nonlinearity of fused silica, where the waveguide array is written. Normalizations for all parameters are provided in \textbf{Methods}.

While the waveguide centers $(x_{n,m},\,y_{n,m})$ far from the dislocation correspond to the nodes of the ideal square lattice, near the dislocation their positions deviate from perfect periodicity. To model this distortion, we introduce a smooth shift in the $x$-coordinates of the waveguides governed by the hyperbolic tangent function, $x \sim \tanh\left(m - \delta_y/d\right)$, where $m$ is an integer (see \textbf{Methods} for the exact expressions for waveguide coordinates), while the parameter $\delta_y$ determines the $y$-position of the inflection point of the $\tanh$ function (with respect to horizontal layers of the structure). The $y$-spacing between layers of waveguides does not change and is equal to $d$. The resulting positions of waveguide layers are depicted as magenta dash-dotted lines in Figs.~\ref{fig1}(a) and~(b) for the cases of two and three merging layers, respectively. We have found that, namely, the shift parameter $\delta_y$ strongly affects the localization properties of the eigenmodes emerging on the dislocation. Other parameters, such as the width of the transition region in the $\tanh$ profile, weakly impact localization as compared to $\delta_y$, so we use the latter as a primary control parameter for tuning our arrays. In Figs.~\ref{fig1}(a) and~(b) with microphotographs of the waveguide arrays, red horizontal lines indicate vertical shifts associated with different $\delta_y$ values. This shift $\delta_y$ is larger in the left panels, and this usually leads to a stronger localization of dislocation modes. Notice that $\delta_y$ can be easily controlled during fs-laser inscription, affording remarkable control over localization properties through tunable dislocation geometries.

First, by omitting cubic nonlinearity in Eq.~(\ref{NLS}), we obtain the linear eigenmodes of our arrays in the form $\psi(x,y,z) = u(x,y) e^{i b z}$, where $u$ is a real-valued function describing the mode profile, and $b$ is the propagation constant. The eigenvalue spectra as functions of the shift parameter $\delta_y$ are shown in Fig.~\ref{fig1}(c) for the array with two merging layers and in Fig.~\ref{fig1}(d) for the array with three merging layers. The gray-to-red color scale indicates in addition the form-factor $\chi=\left[\iint|u|^4\, dxdy\right]^{1/2}/\iint|u|^2\, dxdy$ of all eigenmodes, which quantifies the degree of their localization: the larger is the value of $\chi$, the stronger the localization of the mode. For both structures, at small values of $\delta_y$ only a continuous band of delocalized states. As $\delta_y$ increases, two modes progressively separate from the band, with one emerging above and the other below the band. Namely, these modes represent localized dislocation modes, with both their spectral separation from the band and their localization degree (quantified $\chi$) increasing with $\delta_y$, while all other states within the band remain delocalized. A remarkable distinction between the linear spectra in Figs.~\ref{fig1}(c) and~(d) lies in the onset of localization: for the first dislocation type, localized modes appear only when the shift $\delta_y$ becomes positive, while in the second case, localization already occurs at negative $\delta_y$ and becomes more pronounced as $\delta_y$ increases.

Representative profiles of localized eigenmodes on dislocation are shown in Figs.~\ref{fig1}(e) and~(f), where green color indicates the zero field, red color denotes the positive field values, and blue corresponds to the negative field values. For the structure with two merging layers [Fig.~\ref{fig1}(e)], the upper branch (magenta stars 1–3) corresponds to the symmetric or ``in-phase'' modes with intensity maxima in two waveguides at the dislocation. In contrast, the lower branch (magenta stars 5 and 6) corresponds to antisymmetric or ``out-of-phase'' modes located on the same waveguides. For comparison, a delocalized mode within the band corresponding to magenta star 4 is also shown. For the structure with three merging waveguide layers [Fig.~\ref{fig1}(f)], the dislocation modes exhibit a different structure. The upper branch (magenta diamonds 1–3) corresponds to in-phase modes that now reside on three waveguides aligned horizontally near the dislocation. The lower branch (magenta diamonds 5 and 6) corresponds to out-of-phase modes, appearing on the same waveguides. A representative delocalized in-band state is shown as magenta diamond 4. To highlight the generality of our findings, we also analyze the spectrum and eigenmodes for an edge dislocation with a terminated lattice layer, as well as for an edge dislocation featuring four merging layers (see \textbf{Supplemental Materials}). This configuration also supports dislocation modes with tunable localization. Thus, unlike all previously studied microwave-range structures, our system, designed for optical frequencies, offers multiple control parameters that enable precise tuning of the symmetry and the degree of localization of modes bound to real-space topological lattice defects. These properties of linear modes are expected to strongly affect the properties of new entities -- dislocation solitons that arise in a strongly nonlinear regime.

%%%%%%%%%%%%%%%%%%%%%%%%%%%%%%%%%%%%%%%%%%%%%%%%%%%%%%%%%%%%%%%%
\begin{figure*}[t]
\centering
\includegraphics[width=1.00\linewidth]{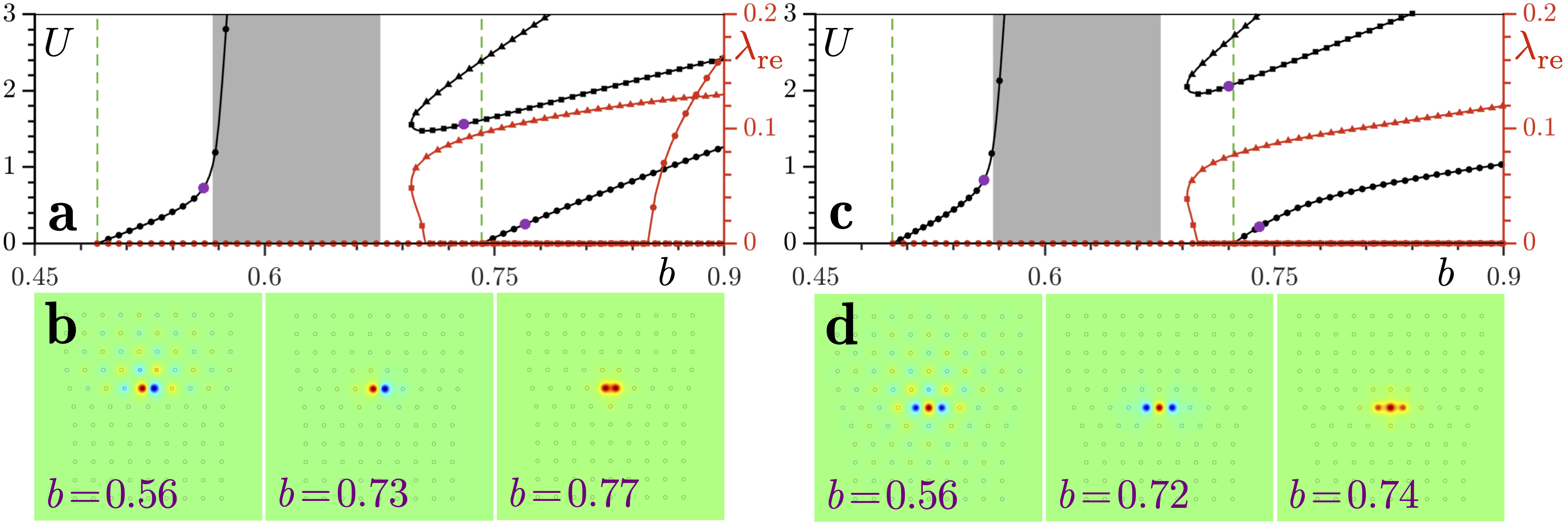}
\caption{{\bf Families of solitons on dislocation.} Power $U$ versus propagation constant $b$ for families of solitons (black lines) in arrays with dislocations with two {\bf (a)} and three {\bf (c)} merging layers. Panel {\bf (a)} corresponds to $\delta_y=+3.2$, while panel {\bf (c)} corresponds to $\delta_y=-1.6$. Gray region corresponds to the bulk band. Vertical dashed lines indicate the eigenvalue of the linear dislocation states from which thresholdless soliton families bifurcate. Red lines in {\bf (a)} and {\bf (c)} show perturbation growth rates for all depicted soliton families (the symbols in $\lambda_{\text{re}}(b)$ dependencies correspond to symbols on the respective soliton families). Profiles of solitons corresponding to violet dots in {\bf (a)} and {\bf (c)} are shown in {\bf (b)} and {\bf (d)}, respectively. Black circles in {\bf (b)} and {\bf (d)} indicate positions of the waveguides in the array.}
\label{fig2}
\end{figure*}
%%%%%%%%%%%%%%%%%%%%%%%%%%%%%%%%%%%%%%%%%%%%%%%%%%%%%%%%%%%%%%%%

{To investigate the bifurcation of dislocation soliton families from linear modes, we now consider Eq.~(\ref{NLS}) with cubic nonlinearity included. We will demonstrate that such solitons can also be excited experimentally. First, however, we obtain these soliton solutions numerically and analyze their stability.} Such solitons may emerge (bifurcate) from linear dislocation modes described above due to the action of focusing nonlinearity. {These nonlinear localized states were computed from Eq.~(\ref{NLS}) using the substitution $\psi(x,y,z) = w(x,y) e^{i b z}$ , where $w$ is a real function and $b$ is the nonlinear propagation constant that determines the soliton power $U = \iint |w|^2\,dxdy$. Soliton profiles were found using Newton iterations method.} The power curves $U(b)$ for dislocation solitons are shown in black in Fig.~\ref{fig2}(a) for the dislocation with two merging layers and $\delta_y = 3.2$ and in Fig.~\ref{fig2}(c) for the dislocation with three merging layers and $\delta_y = -1.6$. For these parameters, the linear spectrum supports well-localized dislocation modes with either in-phase or out-of-phase spots and solitons bifurcating from them inherit their representative internal structure [see examples of $w$ distributions in Figs.~\ref{fig2}(b) and \ref{fig2}(d)]. {In addition, the fact that dislocation solitons bifurcate from linear modes follows from observation that soliton power $U$ vanishes exactly at propagation constant value $b$ corresponding to linear eigenmode with a given symmetry (recall, that in uniform periodic lattices, where eigenmodes are extended Bloch waves, solitons can exist only above power threshold $U$ depending on lattice depth -- hence, the presence of the point where $U\to 0$ is a direct indication of the existence of linear mode on dislocation).} For the in-phase solitons, the bifurcation point (where the power $U$ vanishes) is located above the linear band (gray region) and is marked by a vertical dashed line for both configurations. As power $U$ increases, the propagation constant $b$ shifts upwards into the semi-infinite gap, resulting in progressively increasing nonlinear localization. In contrast, the out-of-phase solitons bifurcate from a point below the band, also indicated by a vertical dashed line. Since the propagation constant $b$ of such states also increases with power $U$, it eventually enters into the band that results in broadening and eventual delocalization of solitons. Importantly, in both structures, solitons can exist without a power threshold. This thresholdless behavior sharply contrasts with properties of solitons in purely periodic arrays, where they exist only above the power threshold. Notice that slightly above the band, multiple soliton branches with threshold emerge due to the hybridization of the solitons with extended bulk modes (see black curves with square and triangle symbols). As $\delta_y$ decreases, the associated linear dislocation modes gradually delocalize, eventually becoming extended states. This transition profoundly affects the properties of the solitons, introducing a nonzero power threshold when no localized linear modes are present in the spectrum. Therefore, by continuously tuning $\delta_y$, i.e., adjusting the waveguide positions near the dislocation, one can directly control one of the key nonlinear features: the soliton excitation threshold.

To assess the experimental feasibility of excitation of dislocation solitons, we conducted a linear stability analysis for corresponding soliton families (see \textbf{Methods}). Alongside the soliton families $U(b)$ (black curves) in Figs.~\ref{fig2}(a) and~(c), we plot the corresponding maximal perturbation growth rate $\lambda_{\mathrm{re}}$ (red curves), where symbols in the $\lambda_{\mathrm{re}}$ plots match those of the respective soliton families $U(b)$. A soliton is linearly stable if $\lambda_{\mathrm{re}} \leq 0$, and unstable if $\lambda_{\mathrm{re}} > 0$, since in the latter case, some perturbations may grow as $\sim e^{\lambda_{\mathrm{re}} z}$ and eventually destabilize the solution. The stability properties differ for the two types of dislocations. For the structure with three merging layers [Fig.~\ref{fig2}(c)], the in-phase thresholdless soliton branch remains entirely stable. In contrast, for the structure with two merging layers [Fig.~\ref{fig2}(a)], in-phase thresholdless solitons are only stable near the bifurcation point, but become unstable with increase of power [see red curve with circles in Fig. \ref{fig2}(b) indicating on the appearance of perturbations with positive $\lambda_{\mathrm{re}}$, i.e. on the onset of instability]. Remarkably, out-of-phase solitons below the band are found to be stable in both structures. For branches with a threshold above the band, only the lower portion remains stable, while the upper part is entirely unstable. To illustrate the generality of our findings, we also present soliton families and analyze their stability for edge dislocations featuring four merging layers and one terminating layer in the \textbf{Supplemental Materials}. Interestingly, despite the overall expansion of soliton profiles in structures with a larger number of merging layers, such solitons remain stable, especially out-of-phase branches. This analysis of nonlinear modes allows us to identify the parameter regimes where stable solitons can be excited experimentally.

%%%%%%%%%%%%%%%%%%%%%%%%%%%%%%%%%%%%%%%%%%%%%%%%%%%%%%%%%%%%%%%%
\subsection*{Observation of dislocation solitons}\label{sec12}

%%%%%%%%%%%%%%%%%%%%%%%%%%%%%%%%%%%%%%%%%%%%%%%%%%%%%%%%%%%%%%%%
\begin{figure*}[t]
\centering
\includegraphics[width=1.00\linewidth]{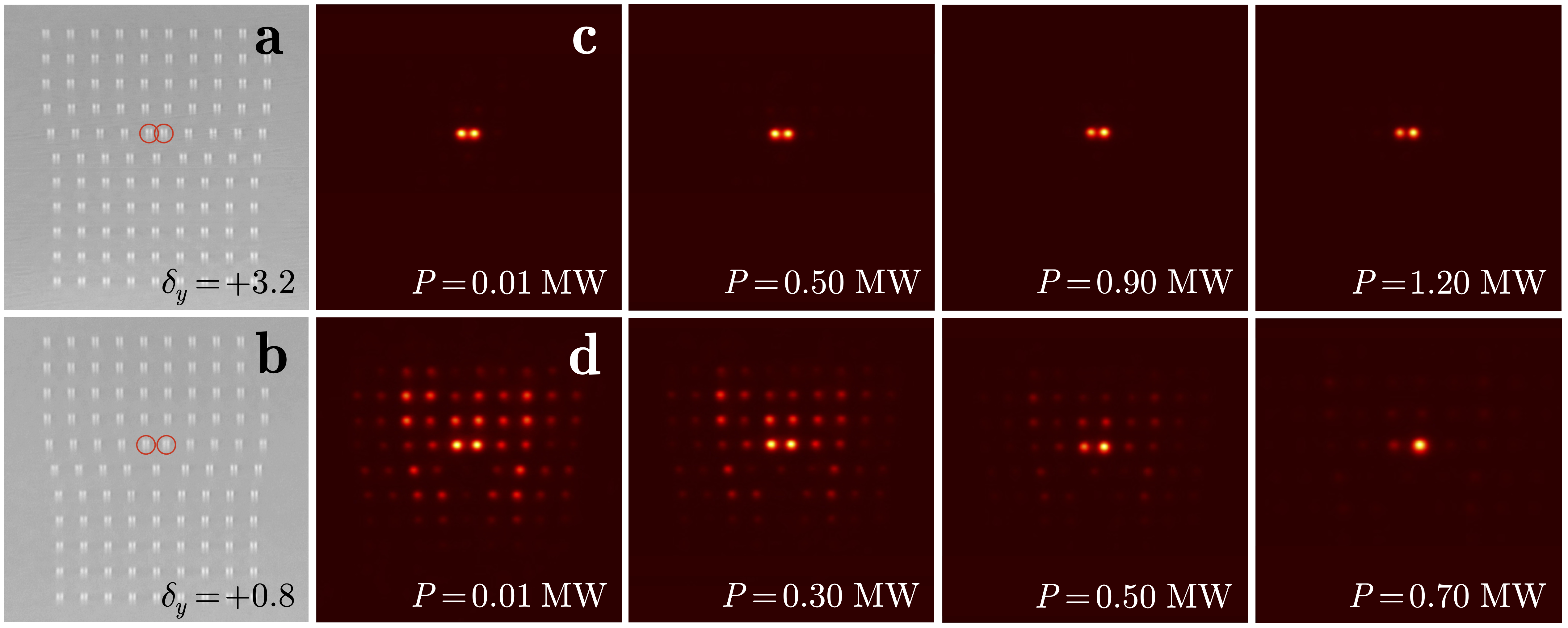}
\caption{{\bf Excitation of the in-phase dislocation solitons.} Microphotographs of the waveguide arrays with dislocation (two merging layers) at $\delta_y=3.2$ {\bf (a)} and $\delta_y=0.8$ {\bf (b)}. Red circles indicate the waveguides that were excited with in-phase beams. Output intensity distributions for different input peak powers $P$ in arrays with $\delta_y=3.2$ {\bf (c)} and $\delta_y=0.8$ {\bf (d)}.}
\label{fig3}
\end{figure*}
%%%%%%%%%%%%%%%%%%%%%%%%%%%%%%%%%%%%%%%%%%%%%%%%%%%%%%%%%%%%%%%%

%%%%%%%%%%%%%%%%%%%%%%%%%%%%%%%%%%%%%%%%%%%%%%%%%%%%%%%%%%%%%%%%
\begin{figure*}[t]
\centering
\includegraphics[width=1.00\linewidth]{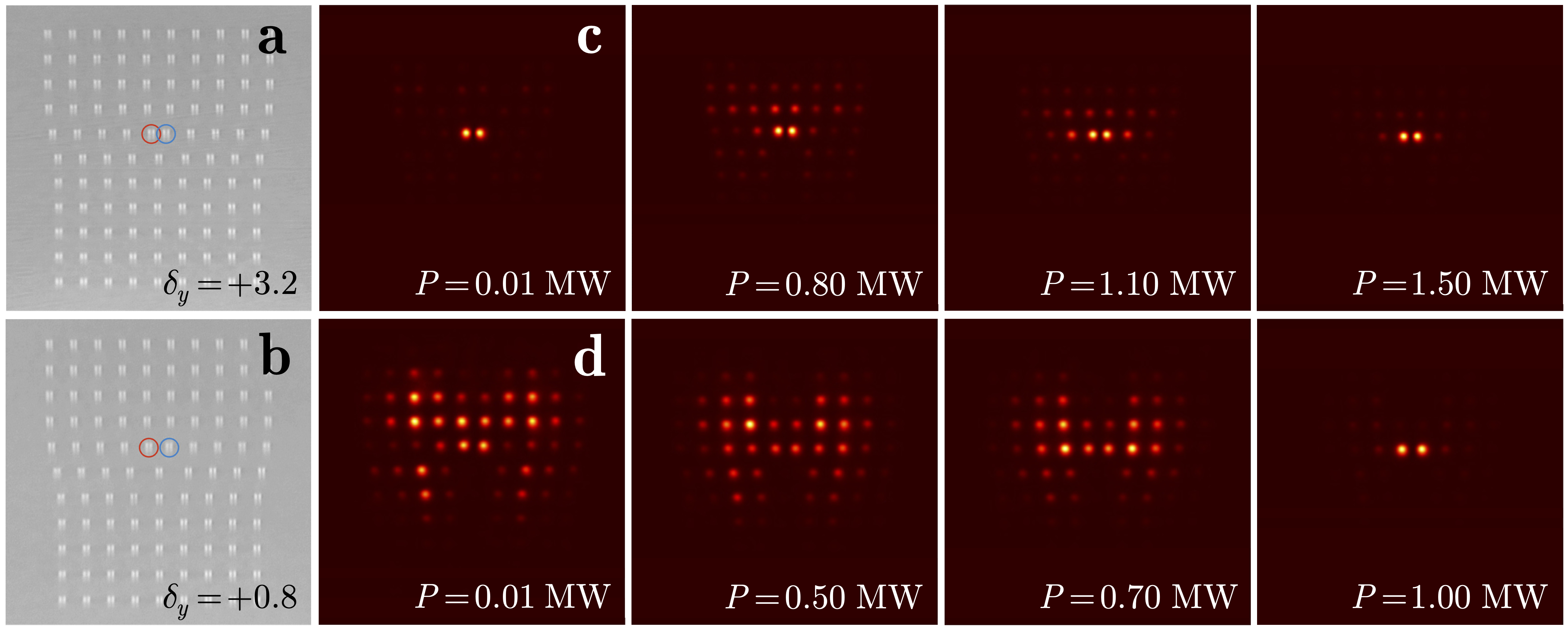}
\caption{{\bf Excitation of the out-of-phase dislocation solitons.} Microphotographs of the waveguide arrays with dislocation (two merging layers) at $\delta_y=3.2$ {\bf (a)} and $\delta_y=0.8$ {\bf (b)}. Red and blue circles indicate the waveguides that were excited with out-of-phase beams. Output intensity distributions for different input peak powers $P$ in arrays with $\delta_y=3.2$ {\bf (c)} and $\delta_y=0.8$ {\bf (d)}.}
\label{fig4}
\end{figure*}
%%%%%%%%%%%%%%%%%%%%%%%%%%%%%%%%%%%%%%%%%%%%%%%%%%%%%%%%%%%%%%%%

%%%%%%%%%%%%%%%%%%%%%%%%%%%%%%%%%%%%%%%%%%%%%%%%%%%%%%%%%%%%%%%%
\begin{figure*}[t]
\centering
\includegraphics[width=1.00\linewidth]{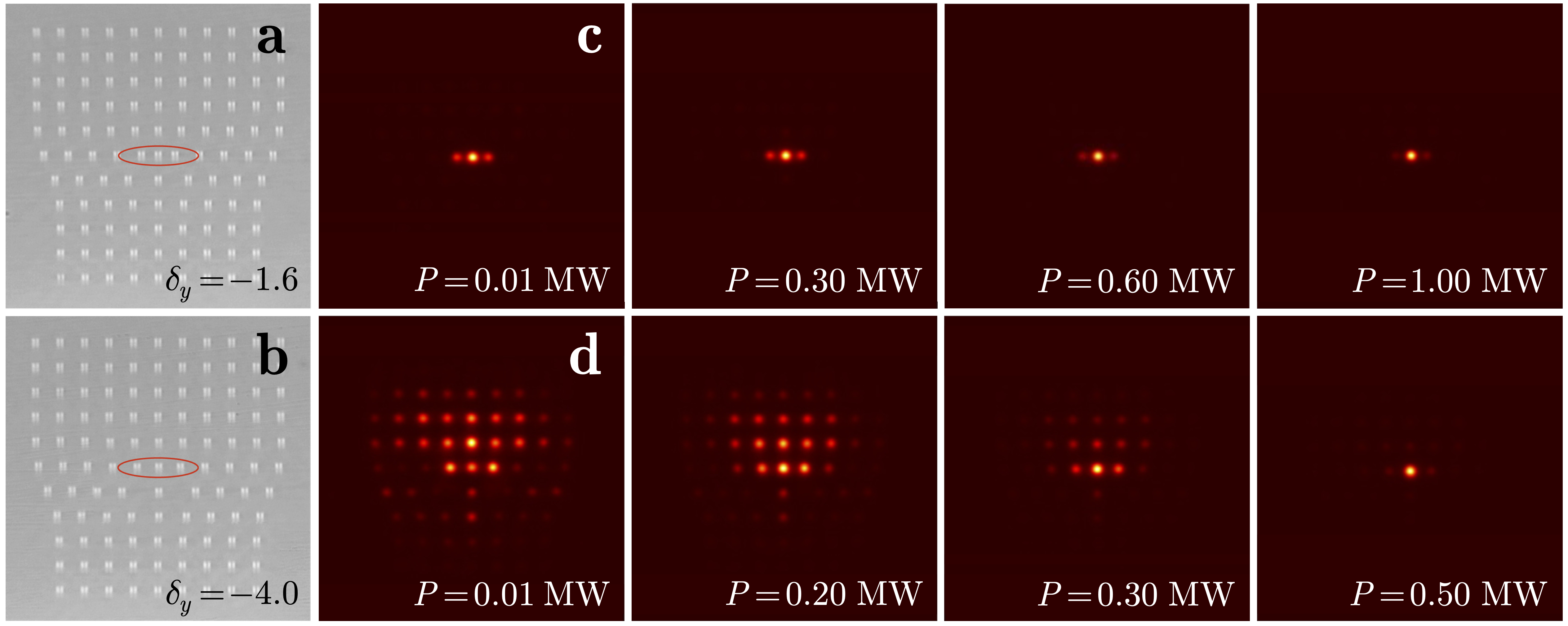}
\caption{{\bf Excitation of the in-phase dislocation solitons on three merging layers.} Microphotographs of the waveguide arrays with dislocation at $\delta_y=-1.6$ {\bf (a)} and $\delta_y=-4.0$ {\bf (b)}, with red circles indicating the elliptical excitation of several central waveguides. Output intensity distributions for different input peak powers $P$ of elliptical beam in arrays $\delta_y=-1.6$ {\bf (c)} and $\delta_y=-4.0$ {\bf (d)}.}
\label{fig5}
\end{figure*}
%%%%%%%%%%%%%%%%%%%%%%%%%%%%%%%%%%%%%%%%%%%%%%%%%%%%%%%%%%%%%%%%

One of the key features of our system is the coexistence of the in-phase and out-of-phase edge solitons, which possess qualitatively distinct intensity and phase profiles. This difference enables the selective excitation of solitons of different types using tailored input beams. To experimentally excite in-phase and out-of-phase solitons in the array with two merging layers, we used two input beams with $0$ or $\pi$ phase difference launched into two waveguides located near the merger of two layers to achieve maximal overlap with the target states [Fig.~\ref{fig2}(b)]. The input beam was derived from a Ti:sapphire laser system operating at a central wavelength of $800$~nm, delivering $1.5$~ps pulses at a repetition rate of $1$~kHz and variable pulse energy $E$, allowing peak powers $P$ up to several megawatts. At such power levels, the Kerr nonlinearity of fused silica starts strongly affecting light propagation. To create two mutually coherent beams with tunable relative phase that are focused into two waveguides, we used a Michelson interferometer (see \textbf{Methods}). Red circles in Figs.~\ref{fig3}(a) and~(b) indicate the excited waveguides with in-phase beams in the array with dislocation at $\delta_y = 3.2$ and $\delta_y = 0.8$, respectively. Corresponding output intensity distributions at different power levels $P$ after the propagation in ${10\,\textrm{cm}}$-long fused silica sample are shown in Figs.~\ref{fig3}(c) and~(d). At $\delta_y = 3.2$, the linear spectrum contains a well-localized in-phase dislocation mode, which is efficiently excited at low powers, as evidenced by the calculated initial modal weight distributions (see \textbf{Supplemental Materials}) and strongly confined output patterns. As the power increases, the in-phase dislocation soliton forms that remains stable within a broad range of input powers, but it starts showing signatures of instability around $P \approx 0.9$~MW, seen as asymmetry of spots in two excited channels (in full agreement with linear stability analysis). In contrast, for $\delta_y = 0.8$, no well-localized linear mode exists and the low-power input excites multiple linear modes simultaneously, the beat of which manifests as a rapid diffraction of the beam into the bulk of the sample. At higher powers, however, nonlinear localization gradually develops, and already at $P \approx 0.5$~MW, a well-confined output pattern appears, which indicates soliton formation, but above the power threshold. Further power increase leads to the onset of instability that may result in the concentration of light in only one of the waveguides. Numerical simulations using similar input conditions, presented in \textbf{Supplemental Materials}, reproduce the experimental intensity patterns, further confirming our observations. Notice that the dependence of the output form-factor on input power $\chi(U)$ illustrating details of qualitatively different excitation dynamics for two considered values of shift $\delta_y$ is also presented in \textbf{Supplemental Materials}.

The excitation of the out-of-phase state in the same arrays reveals a distinct and more complex behavior. Figures~\ref{fig4}(a) and~(b) indicate the excited with out-of-phase beams waveguides (red and blue circles) in structures with $\delta_y = 3.2$ and $\delta_y = 0.8$, respectively. For $\delta_y = 3.2$, the array supports a well-localized out-of-phase linear dislocation mode. Accordingly, we observe a strong localization of the output intensity distributions at low powers. As the input power increases, the output pattern gradually spreads, indicating coupling to extended bulk modes because nonlinearity drives this out-of-phase state into the band of delocalized states. Representative output profiles at intermediate powers ($P = 0.8$ and $1.1~\textrm{MW}$) are shown in Fig.~\ref{fig4}(c). With further increase of power, a transition to a strongly localized state occurs, and an out-of-phase soliton emerging from the semi-infinite gap is formed. No signatures of instability are observed in this regime. In contrast, for $\delta_y = 0.8$, when the array lacks localized modes, low-power excitations lead to strong diffraction. Even at moderate powers, the light remains delocalized, again suggesting coupling to extended modes from the band. At sufficiently high power, the abrupt transition to localization occurs, indicating the formation of an out-of-phase soliton from the semi-infinite gap (existing above the power threshold). This nonlinear state remains robust, with no signs of instability, in agreement with predictions of linear stability analysis. Corresponding numerical simulations of dynamical excitation of the out-of-phase dislocation solitons are also presented in \textbf{Supplemental Materials}.

We have also observed the excitation of in-phase solitons in dislocation arrays with three merging layers. To efficiently excite corresponding states, we used a wide elliptical input beam covering the three waveguides nearest to the dislocation center, as indicated by the red ellipses in Figs.~\ref{fig5}(a) and~(b). The input beam was shaped using a cylindrical telescope, which produced an elliptical spot at the input facet with a full width at half maximum (FWHM) of approximately ${43.8\,\mu\textrm{m}}$ along the $x$-axis and ${12.9\,\mu\textrm{m}}$ along the $y$-axis (see \textbf{Methods}). According to the modal weight analysis (see \textbf{Supplemental Materials}), this beam preferentially excites the dislocation mode when the latter is sufficiently well localized, while in the absence of such modes in the linear spectrum the excitation of multiple extended modes and subsequent diffraction occur. Figures~\ref{fig5}(a) and~(c) illustrate output intensity distributions for the case $\delta_y = -1.6$, where localized dislocation mode exists in the linear spectrum. Here, even low-power excitation leads to strong spatial confinement of light. In contrast, for $\delta_y = -4.0$ [Figs.~\ref{fig5}(b) and~(d)], where no localized linear modes are present, the same excitation results in pronounced diffraction at low powers. In both cases, increasing the input power enhances localization due to the focusing nonlinearity. For $\delta_y = -1.6$, this results in gradual contraction of light practically into a single central waveguide already at peak power $P = 1~\textrm{MW}$. In contrast, at $\delta_y = -4.0$ the transition to localization is observed at a power level of approximately $0.5~\textrm{MW}$. The simulation results for this case are presented in \textbf{Supplemental Materials}, alongside experimental and numerical data for the edge dislocation of different types, containing termination of the layer. In this complementary case, we also observe a nontrivial interplay between nonlinear effects and localization at dislocation, resulting in rich transformations of the output intensity distributions.

%%%%%%%%%%%%%%%%%%%%%%%%%%%%%%%%%%%%%%%%%%%%%%%%%%%%%%%%%%%%%%%%
\section*{Discussion}\label{sec2}

We have presented the first experimental observation of localized linear modes with different internal structures bound to edge dislocations and of dislocation solitons at optical frequencies. Our results reveal a rich interplay between controllable global lattice deformations and nonlinearity, enabling new mechanisms of light localization. The states observed here exhibit several distinctive features stemming from the type of dislocation and local geometry of the array around it. First, each dislocation seems to support two different types of modes with different symmetries that emerge as localized states for nearly the same shift defining the local shape of the array, and as a result, there are also two different types of dislocation solitons available in the system. Second, increasing the number of merging array layers leads to a certain expansion of the modes, but they still remain well-localized. Third, dislocation solitons can form without a power threshold, unlike their counterparts in periodic lattices, and their stability properties differ substantially from those of conventional lattice solitons. Fourth, in contrast to topological edge modes and solitons resulting from a nontrivial topology of spectral bands, dislocation modes tied to real-space topological lattice defects may form in the depth of the structure.

The presence of robust and nonlinearity-tunable light trapping on engineered topological dislocations in photonic systems implies concrete technological applications, since such structures open up unique opportunities for the creation of fundamentally new optical devices with enhanced properties. The study of dislocations and light trapping on them is motivated not only by the fundamental interest in topological photonics, but also by the practical need to overcome the limitations of traditional photonics, such as confinement limit, complexity of integration, and sensitivity of trapping to defects. Dislocations appear to be a powerful tool for creating devices with characteristics that cannot be achieved by traditional methods. For instance, topological dislocations could serve as reconfigurable waveguides or logic gates in integrated photonic circuits, enabling novel schemes for photonic routing and switching. {Such architectures can be constructed from pairs or sequences of closely spaced dislocations, with reconfigurability enabled by the controlled mode localization demonstrated in this work. Incorporating nonlinearity adds an additional degree of freedom, allowing power-dependent switching and further tunability of light routing. By combining different types of dislocations, their complex sequences can be engineered along which light can travel, while remaining potentially resistant to disorder. Closely spaced dislocations may also support compound solitons with unique spatial profiles, offering opportunities for designing photonic elements with tailored confinement and propagation characteristics. Beyond switching applications, these results have implications for laser physics and quantum optics. For instance, the existence of strongly localized modes at dislocations could enable topological microlasers with enhanced mode confinement, robustness, and reduced lasing thresholds.}
%Moreover, the fields of laser physics and quantum optics have long pursued ultra-compact, low-loss lasers.
%Our results suggest that strongly localized modes emerging at dislocations could lead to the realization of topological microlasers, benefiting from spatial mode confinement, enhanced robustness, and reduced lasing thresholds.
% Furthermore, the highly localized and controllable nature of dislocation modes makes them promising candidates for quantum memory platforms, where information encoded in spatial light structure can be stored. {{In the nonlinear regime, such localized states may even support topological qubits, analogous to zero modes in Majorana systems.}}
Potential future research directions include exploration of the non-Hermitian and Floquet systems with embedded dislocations, parametric interactions and solitons in non-Kerr nonlinear systems, and quantum effects.

%%%%%%%%%%%%%%%%%%%%%%%%%%%%%%%%%%%%%%%%%%%%%%%%%%%%%%%%%%%%%%%%
%%%%%%%%%%%%%%%%%%%%%%%%%%%%%%%%%%%%%%%%%%%%%%%%%%%%%%%%%%%%%%%%
\section*{Methods}\label{sec3}

\subsection*{Normalization of parameters in theory}\label{sub1}

The formation of solitons is described by the dimensionless nonlinear Schr\"{o}dinger Eq.~(\ref{NLS}). The transverse coordinates are normalized as $x = X/r_0$, $y = Y/r_0$, where $r_0 = 10$~$\mu\text{m}$ is the characteristic transverse scale. The propagation distance is scaled as $z = Z / L_d$, with the diffraction length $L_d = k r_0^2 \approx 1.14$~mm, where $k = 2\pi n / \lambda$ is the wavenumber for the operating wavelength $\lambda = 800$~nm and the background refractive index $n \approx 1.45$ (fused silica). The dimensionless field amplitude $\psi$ is related to the physical electric field $\mathcal{E}$ via $\psi = (k^2 r_0^2 n_2 / n)^{1/2} \mathcal{E}$, where $n_2 \approx 2.7 \times 10^{-20}$~m$^2$/W is the nonlinear refractive index of fused silica. The dimensionless depth of the array is given by $p = k^2 r_0^2 \delta n / n$, where $\delta n$ is the refractive index contrast. The value $p = 3.1$ used in all simulations corresponds to $\delta n \approx 3.5 \times 10^{-4}$. The waveguide widths $w_x = 0.75$, $w_y = 0.65$ correspond to $7.5$ $\mu$m and $6.5$ $\mu$m, respectively. The waveguide spacing $d = 4$ far from the dislocation center corresponds to $40$~$\mu$m. A physical sample length of $10$~cm corresponds to a dimensionless propagation distance $z \approx 88$.

\subsection*{Coordinates of waveguides in arrays with dislocation}\label{sub2}

Although the waveguide centers $(x_{n,m},\,y_{n,m})$ align with an ideal square lattice far away from the dislocation, they exhibit deviations from perfect periodicity near the dislocation center. As described in the main text, we model these deviations by assuming that the shift of the $x$-positions of the waveguides is described by the hyperbolic tangent functions. In the configuration where two array layers merge into one [see Fig.~\ref{fig1}(a)], the waveguide positions are given by
\begin{align}
\label{xnyndislocation1}
x_{n,m} =
\begin{cases}
n d + \frac{d}{4}\left[3 - \tanh\left({m - \frac{\delta_y}{d}}\right)\right], & \text{ for } n \le -1, \\
%n d, & \text{ for } n = 0, \\
n d - \frac{d}{4}\left[3 - \tanh\left({m - \frac{\delta_y}{d}}\right)\right], & \text{ for } n \ge +1,
\end{cases}
\end{align}
%\begin{widetext}
%\begin{align}
%\label{xnyndislocation1}
%\begin{cases}
%x_{n,m} = n d + \frac{d}{4}\left[3 - \tanh\left({m - \frac{\delta_y}{d}}\right)\right],\,y_{n,m} = m d, & \text{ for } n < -1, \,m\in\mathbb{Z},\\
%x_{n,m} = n d + \frac{d}{4}\left[3 - \tanh\left({m - \frac{\delta_y}{d}}\right)\right],\,y_{n,m} = m d, & \text{ for } n = -1, \,m>0,\\
%x_{n,m} = n d,\,y_{n,m} = m d, & \text{ for } n = 0,\, m\le0,\\
%x_{n,m} = n d - \frac{d}{4}\left[1 + \tanh\left({m - \frac{\delta_y}{d}}\right)\right],\,y_{n,m} = m d, & \text{ for } n = +1,\,m>0,\\
%x_{n,m} = n d - \frac{d}{4}\left[1 + \tanh\left({m - \frac{\delta_y}{d}}\right)\right],\,y_{n,m} = m d, & \text{ for } n > +1,\,m\in\mathbb{Z},
%\end{cases}
%\end{align}
%\end{widetext}
while $y_{n,m} = md$. For the structure with three merging layers [see Fig.~\ref{fig1}(b)], the positions of waveguides are given by
\begin{align}
\label{xnyndislocation2}
x_{n,m} =
\begin{cases}
n d + \frac{d}{2}\left[1 - \tanh\left({m - \frac{\delta_y}{d}}\right)\right], & \text{ for } n \le -1, \\
n d, & \text{ for } n = 0, \\
n d - \frac{d}{2}\left[1 - \tanh\left({m - \frac{\delta_y}{d}}\right)\right], & \text{ for } n \ge +1,
\end{cases}
\end{align}
%\begin{widetext}
%\begin{align}
%\label{xnyndislocation1}
%\begin{cases}
%x_{n,m} = n d + \frac{d}{2}\left[1 - \tanh\left({m - \frac{\delta_y}{d}}\right)\right],\,y_{n,m} = m d, & \text{ for } n < -1, \,m\in\mathbb{Z},\\
%x_{n,m} = n d + \frac{d}{2}\left[1 - \tanh\left({m - \frac{\delta_y}{d}}\right)\right],\,y_{n,m} = m d, & \text{ for } n = -1, \,m\ge0,\\
%x_{n,m} = n d,\,y_{n,m} = m d, & \text{ for } n = 0,\, m\in\mathbb{Z},\\
%x_{n,m} = n d - \frac{d}{2}\left[1 - \tanh\left({m - \frac{\delta_y}{d}}\right)\right], & \text{ for } n = +1,\,m\ge0,\\
%x_{n,m} = n d - \frac{d}{2}\left[1 - \tanh\left({m - \frac{\delta_y}{d}}\right)\right],\,y_{n,m} = m d, & \text{ for } n > +1,\,m\in\mathbb{Z},
%\end{cases}
%\end{align}
%\end{widetext}
and again $y_{n,m} = md$. In both cases, $n$ and $m$ are integer numbers. Note that in the cases, where the coordinates of waveguide centers coincide, only a single waveguide is retained. The parameter $\delta_y$ sets the $y$-coordinate of the inflection point of the $\tanh$ function, while the vertical spacing between waveguide rows remains constant and equal to $d$.

\subsection*{Stability analysis of solitons}\label{sub3}

To assess the stability of solitons, we performed a linear stability analysis by considering a perturbed solutions of the form $\psi=\left[w(x,y)+u(x,y)e^{\lambda z}+v^*(x,y)e^{\lambda^*z}\right]e^{ibz}$, where $u(x,y)$ and $v(x,y)$ are small perturbations satisfying the conditions $\iint |u|^2\,dxdy$, $\iint |v|^2\,dxdy\ll\iint |w|^2\,dxdy$, and $\lambda = \lambda_{\text{re}} + i \lambda_{\text{im}}$ is the complex perturbation growth rate. Substituting this ansatz into Eq.~(\ref{NLS}) and linearizing it around the stationary soliton solution $w(x,y)$ yields the linear eigenvalue problem
\begin{equation} \label{stability}
    \begin{split}
    \lambda u&=+i\left[(1/2)\Delta u+\mathcal{R}u-bu+2|w|^2u+w^2v\right], \\
    \lambda v&=-i\left[(1/2)\Delta v+\mathcal{R}v-bv+2|w|^2v+w^2u\right],
    \end{split}
\end{equation}
where $\Delta = \partial^2 / \partial x^2 + \partial^2 / \partial y^2$ is the transverse Laplacian. Eqs.~(\ref{stability}) were solved numerically. A soliton is linearly stable if $\lambda_{\text{re}} \leq 0$ for all eigenvalues, and unstable if there exists at least one eigenvalue with $\lambda_{\text{re}} > 0$. In Fig.~\ref{fig2}, we display only the largest values of the perturbation growth rate $\max\left[\lambda_{\text{re}}\right]$ for each propagation constant $b$ and each soliton branch.

\subsection*{Fs-laser inscription of dislocation arrays}\label{sub4}

Waveguide arrays with dislocations were fabricated in a ${10\,\textrm{cm}}$-long fused silica substrate (JGS1) using femtosecond laser direct writing. A circularly polarized laser beam with central wavelength ${515\,\textrm{nm}}$, pulse duration ${230\,\textrm{fs}}$, repetition rate ${1\,\textrm{MHz}}$, and pulse energy ${270\,\textrm{nJ}}$ was tightly focused into the substrate by an aspheric lens ($\textrm{NA} = 0.4$), enabling writing of waveguides with practically identical refractive index contrast over a depth range of ${600\,\mu\textrm{m}}$. The sample was translated with respect to the laser beam using a high-precision motion control system (AeroTech). We employed a multiscan writing strategy~\cite{Skryabin-2024} to reduce the intrinsic ellipticity of the waveguides and to guarantee isotropy of coupling. Each waveguide was composed of six adjacent tracks separated by ${1.6\,\mu\textrm{m}}$, resulting in a rectangular cross-section and nearly circular eigenmode with aspect ratio $\approx 0.98$. The scan speed for each track was set to ${30\,\textrm{mm/s}}$, resulting in an effective writing speed of ${5\,\textrm{mm/s}}$ per waveguide. The resulting waveguides exhibited low propagation losses of ${0.1\,\textrm{dB/cm}}$ at the operational wavelength of ${800\,\textrm{nm}}$.

\subsection*{Experimental excitation conditions}\label{sub5}

To excite solitons, we used a Ti:Sapphire laser system (Spitfire Pro, Spectra Physics) with a repetition rate of $1~\textrm{kHz}$ and a pulse duration of $40$~fs at a central wavelength of $800$~nm. To minimize the effects connected with pulse shape transformation upon propagation inside the sample, initially short pulses with a wide spectrum were narrowed using a $10$~nm interference filter and temporally stretched to $1.5$~ps via a built-in grating compressor. After an active laser beam stabilization system (BPS, Avesta) and an attenuator, the beam was focused onto the input face of the sample by an aspherical lens with a focal length of $100$~mm, which provided the calculated overlap integral with the waveguide mode of $0.85$. The sample was mounted on a 6-axis high precision nanopositioner (I6000 6-Axis XYZ/RYP, Luminos).

For controllable excitation of the in-phase and out-of-phase dislocation solitons, we used a Michelson interferometer, where the phase between the two beams was controlled by precise rotation of a pair of compensation plates in one of the arms without altering the direction of the radiation. To measure the intensity distribution at the output of the sample with inscribed waveguides, we used a scientific CMOS camera (Kiralux 12.3~Mp, Thorlabs). For the experiment with excitation of three waveguides by elliptical beam, in front of the focusing lens the beam was shaped using a 1:3 telescope consisting of two cylindrical lenses with focal lengths of $300$ and $100$~mm, respectively.

%%%%%%%%%%%%%%%%%%%%%%%%%%%%%%%%%%%%%%%%%%%%%%%%%%%%%%%%%%%%%%%%%%
\begin{acknowledgments}
%\textbf{Acknowledgments.}
%We are grateful to ... for useful discussions.
\end{acknowledgments}
%%%%%%%%%%%%%%%%%%%%%%%%%%%%%%%%%%%%%%%%%%%%%%%%%%%%%%%%%%%%%%%%%%

~

\textbf{Funding.}
S.K.I. has received funding from the European Union through the Program Fondo Social Europeo Plus 2021-2027(FSE+) of the Valencian Community (Generalitat Valenciana CIAPOS/2023/329). This work was supported by the Russian Science Foundation (grant 24-12-00167) and partially by the research project FFUU-2024-0003 of the Institute of Spectroscopy of the Russian Academy of Sciences. S.A.Z. acknowledges support by the Foundation for the Advancement of Theoretical Physics and Mathematics ``BASIS'' (22-2-2-26-1).

%%%%%%%%%%%%%%%%%%%%%%%%%%%%%%%%%%%%%%%%%%%%%%%%%%%%%%%%%%%%%%%%%%

\textbf{Data availability.}
Data supporting conclusions of this work are included within the Article and its Supplementary Information. All other raw data that support the findings of this study are available from the corresponding author on reasonable request.

\textbf{Authors contributions.}
All authors contributed significantly to this work.

\textbf{Competing interests.}
The authors declare no conflicts of interest.

%\bibliographystyle{apsrev4-2}
%\bibliography{refs}

%%%%%%%%%%%%%%%%%%%%%%%%%%%%%%%%%%%%%%%%%%%%%%%%%%%%%%%%%%%%%%%%%%

\end{document}